\documentclass[prd,superscriptaddress,onecolumn,showpacs,11pt,msmath,preprintnumbers,showkeys]{revtex4}
\usepackage{wrapfig,rotating}
\usepackage{graphicx,epsfig,color}

\makeatletter

\newenvironment{figurehere}
{\def\@captype{figure}}
{}
\makeatother

\usepackage{graphicx}% Include figure files
\usepackage{dcolumn}% Align table columns on decimal point
\usepackage{bm}% bold math

\def\beq{\begin{equation}}
\def\eeq{\end{equation}}
\def\beeq{\begin{eqnarray}}
\def\eeeq{\end{eqnarray}}

\newcommand \Pomeron {I\!\!P}
\newcommand \R {I\!\!R}

\def\effs {$\sigma_{\textrm{\tiny eff}}\,$}
\def\2GPD{$_2\mbox{GPD}$}

\def\12{$1\otimes 2$}
\def\22{$2 \otimes 2$}

\def\Qsep{Q_{\mbox{\rm\scriptsize sep}}}
\def\Qsep2{Q^2_{\mbox{\rm\scriptsize sep}}}

\begin{document}
 \title{%On the soft unfactorizable correlations in the central kinematics in MPI.}
 Interplay of soft and perturbative correlations in multiparton interactions at central rapidities}
 \author{B.\ Blok$^{1}$,
M. Strikman$^2$ \\[2mm] \normalsize $^1$ Department of Physics, Technion -- Israel Institute of Technology,
 Haifa, Israel\\
 \normalsize $^2$Physics Department, Pennsylvania State University, University Park,USA}
 \begin{abstract}
We study the role of soft/nonperturbative correlations in the  multi parton interactions in the central kinematics
relevant for double parton scattering (DPS)  and underlying event (UE)
measurements at ATLAS and CMS. We show that the effect of soft correlations is negligible for DPS
regime (typical transverse momenta larger than 10-20 GeV), but may be important for UE (several GeV scale). The characteristic scale where soft correlations become important increases with decrease of $x$ (energy increase) leading to approximately constant \effs at small x.

 \end{abstract}

   \maketitle
 \thispagestyle{empty}

 \vfill

\section{Introduction.}
\par It is widely realized now that hard {\em Multiple Parton Interactions}\/ (MPI)
occur with a probability of the order one in  typical inelastic LHC proton-proton $pp$ collisions and hence   play an important role in the description of inelastic  $pp$ collisions.
MPI were first introduced in the eighties
 \cite{TreleaniPaver82,mufti} and in the last decade became a subject of a number of the theoretical studies,  see e.g. \cite{stirling,BDFS1,Diehl2,stirling1,BDFS2,Diehl,BDFS3,BDFS4,Gauntnew,BO,Gauntdiehl,mpi2014,mpi2015,BG1,BG2} and references therein.
\par Also,
in the past several years  a number of Double Parton Scattering (DPS) measurements in different channels in the central rapidity  kinematics were carried out
\cite{tevatron1,tevatron2,tevatron3,cms1,atlas,cms2}, while many Monte Carlo (MC)  event generators now incorporate MPIs.
\par It was pointed out starting with \cite{Frankfurt2,BDFS1} that the rate of DPS
can be calculated through the integral over the generalized parton  distributions (GPD) under assumption that partons are not correlated and double GPD is simply a product of single GPDs. Information about  GPDs at small x is available from the analyses of the HERA
\cite{Frankfurt,Frankfurt2}. Based on these analyses it was demonstrated that  the   uncorrelated model predicts the DPS rates which are
a factor $\sim$ 2
 too low to explain the data, hence indicating presence of significant parton - parton correlations.

It was pointed out in
\cite{BDFS4,Gauntnew,BG1,BG2,BO} that correlations generated in the course of the DGLAP evolution --  \12 mechanism -- explain the DPS
 rates in the central rapidity region  \cite{BDFS4,Gauntnew,BG1,BG2,BO}
 provided the starting scale for the QCD evolution -- $Q_0^2= 0.5 \div 1 {\rm GeV}^2$ is chosen.
\par The role of \12 mechanism decreases however once we move to smaller $x$ \cite{BDFS4}. On the other hand as it was explained in \cite{BDFS3} with the decrease of $x$ the relative importance of soft correlations in the nucleon increases. These
correlations lead to a  non-factorizable initial conditions for the evolution of double GPD at initial scale $Q_0$. In our recent
paper \cite{BS1} we presented a simple way to take into account these correlations for the forward kinematics recently studied by LHCb which corresponds to  $x \sim 10^{-4}$. The method is based on  the connection between
the correlation contribution to the MPI and inelastic diffraction \cite{BS1}. We demonstrated that taking into account the mean field
\22 contribution, \12 contribution and soft correlations leads to a good description of the DPS data in the  forward LHCb kinematics \cite{Belyaev,LHCb}. (These data are the most accurate data to  date on the DPS and have
the smallest
background from the leading twist processes).
Moreover it was pointed out that the account of the soft correlations leads to a strong reduction of the dependence of the predicted $\sigma_{eff}$ on the incident energy and $p_t$ of a minijet,    and to $\sigma_{eff}$ being approximately constant
for minijets contributing to  the Underlying event (UE) \cite{BS1}.Moreover, including  soft correlations  have led to a significant reduction  of the sensitivity of the results  to the value of the    parameter $Q_0$ separating soft and hard correlations \cite{BDFS4}.
\par Naturally, it will be interesting to check the consequences of the soft correlations model developed in \cite{BS1}
for  the central kinematics at the LHC relevant for CMS and ATLAS experiments to clarify the role of soft correlations in this kinematics.
\par Hence in this paper we will extend the model of \cite{BS1} to the central kinematics, and quantify the role of  the soft correlations in central kinematics. We shall see that
for  large transverse scales ($Q>20$ GeV)
DPS, considered in \cite{BDFS4,Gauntnew,BG1,BG2,BO},
 they constitute a small correction  with dominant correlations originating from the perturbative \12 mechanism. At the same time we find that soft correlations are important   for the few GeV  transverse scales corresponding to UE,  leading to a weak dependence of $\sigma_{eff}$ on  $Q$ for $Q$  of  the order several GeV
 followed by
  a decrease of $\sigma_{eff}$ at $Q > 10-15 $ GeV largely due to the  \12 contribution.
 Also, similar to the case of the forward kinematics inclusion of the soft contribution makes the result less sensitive to the  different initial scales $Q_0$.
\par For simplicity we will limit our analysis to production of four jets in the gluon interactions.

\noindent
The paper is organized as  follows. In section 2  we summarize the results of our previous analyses of the \22,\12 mechanisms in the central kinematics.
In section 3 we discuss the soft correlations, and in section 4 investigate their role numerically. Our conclusions are presented in section 5.

\section{Mean field approximation  and \12 mechanism estimate
of \effs in central kinematic.}
\begin{figure}[htbp]
\begin{center}
\includegraphics[scale=0.3]{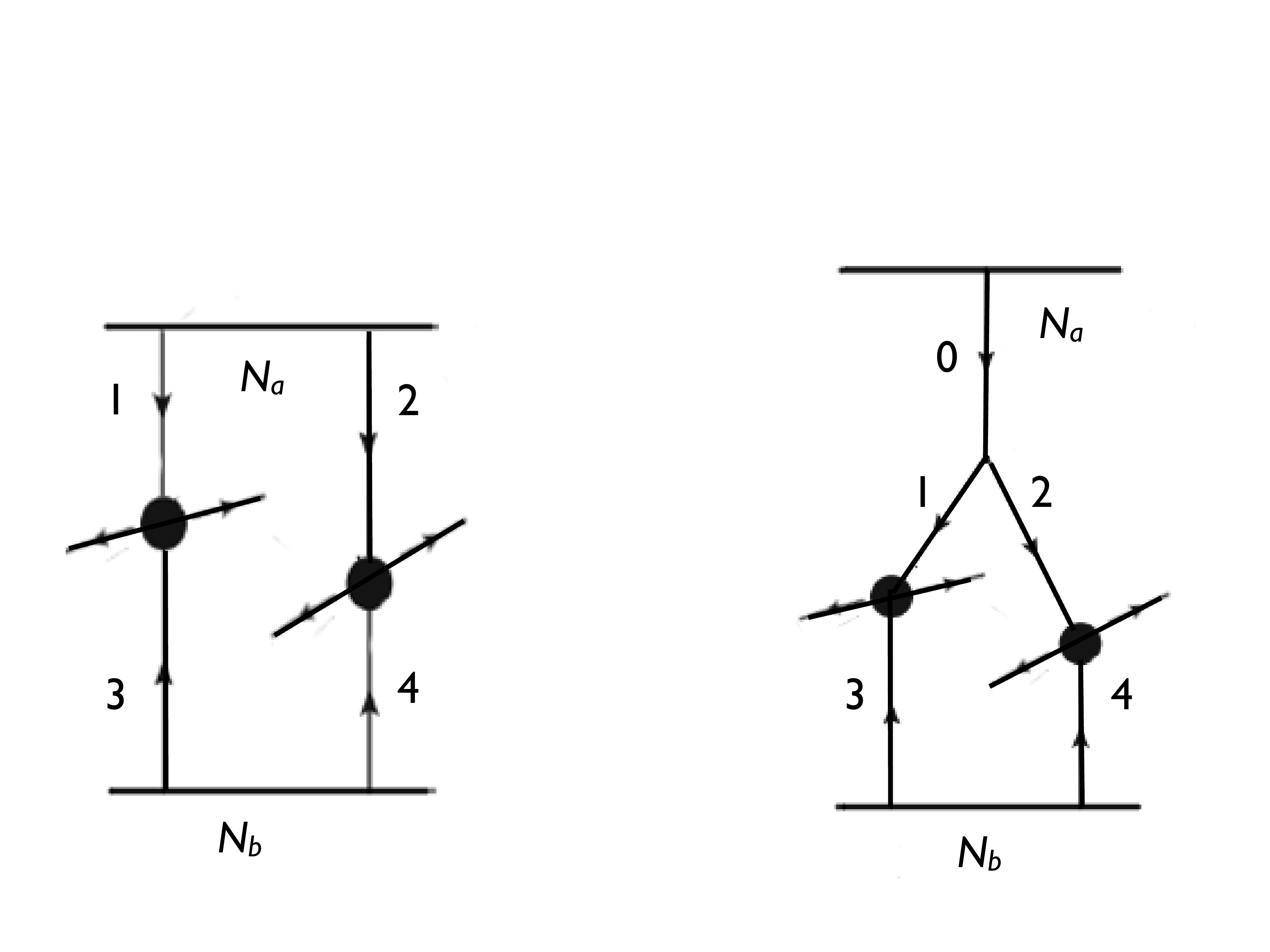}\hspace{3cm}
\label{fig1a}
\caption{Sketch of the two considered DPS mechanisms: \22 (left) and \12 (right) mechanism.}
\end{center}
\end{figure}

If the partons are uncorrelated (the mean field approach)
the
 double  parton GPDs, describing the DPS, are
 given by the product of the single parton GPDs:
\beq
_2D(x_1,x_2,Q_1^2,Q_2^2,\Delta))= _1D(x_1,Q_1^2,\Delta_1) \cdot  _1D(x_2,Q_2^2,\Delta_2),\label{fac}
\eeq
where the one particle GPDs $_1D$ are  known from the analyses \cite{Frankfurt,Frankfurt1} of exclusive $J/\Psi$ photoproduction at HERA.
One can    parameterize GPDs as
\beq
D_1(x,Q^2,\Delta)=D(x,Q^2) F_{2g}(\Delta,x).\label{slon1}
\eeq
Here $D(x,Q^2)$
is the conventional gluon PDF of the nucleon, and $F_{2g}(\Delta,x)$ is the two gluon nucleon form factor.
The effective cross section \effs \, is then given by
\beq
1/\sigma_{eff}=\int \frac{d^2\Delta}{(2\pi)^2}F^4(\Delta)=\frac{1}{2\pi}\frac{1}{B_g(x_1)+B_g(x_2)+B_g(x_3)+B_g(x_4)}
\eeq
Here we use exponential parametrization \cite{Frankfurt1}:
\beq
F_{2g}(\Delta,x)=\exp(-B_g(x)\Delta^2/2),B_g(x)= B_0 + 2K_Q\cdot\log(x_{0}/x)
\label{d1a}
\eeq
 with $x_0\sim 0.0012$, $B_0=4.1$ GeV$^{-2}$ and $K_Q=0.14$ GeV$^{-2}$
(very weak $Q^2$ dependence of $B_g$ is neglected).
Hence we find for the  mean field value of $\sigma_{eff}$\,
at the LHC energies
in
  the central  kinematics  $x_1\sim x_3= \sqrt{4Q_1^2/s}  ,x_2\sim x_4\sim \sqrt{4Q_2^2/s} $,
   $\sigma^{MF}_{eff}$ drops from $\approx$ 43 mb at $Q_i \sim \mbox{2 GeV}$ to
 $\approx$ 37 mb at  $ Q_i \sim  20 GeV$ due to the increase of  $x$  with increase of $Q_i$
 and increase of the transverse area occupied by gluons with decrease of $x$, cf. eq.\ref{d1a}.
  For simplicity we shall consider here the symmetric case $Q_1\sim Q_2$.

\par Presence of the  \12 mechanism (Fig. 1b) in addition to uncorrelated mechanism  of Fig. 1a  leads to the  enhancement of the rate of DPS (increase of
$1/\sigma_{eff}$ ) as compared to its
 mean field value. The \12 mechanism   was suggested in \cite{BDFS1,BDFS2,BDFS3,BDFS4}, where it was demonstrated  that taking into account the pQCD DGLAP ladder splittings leads to a decrease of $\sigma_{eff}$.

We calculate $R_{pQCD}$ - the ratio of the contributions of \12 and mean field mechanisms
 by solving by iterations the evolution equation for $_2GPD$ \cite{BDFS2}.
The numerical
 results
  for the enhancement coefficient $R_{pQCD}$ for \effs
(here $\sigma_{DPS}$ includes \22 and \12 mechanisms)
are summarized in  Fig. 4 below.

 \section{Non-factorized
 contribution
  to $_2D$ at the initial  $Q_0$ scale.}

 \par There is an additional contribution to the  DPS at small $x$ which was first discussed in \cite{BDFS3}, and in more  detail in  \cite{BS1}
 which is related to the soft dynamics.

 It was demonstrated in \cite{BS1} that soft dynamics leads to positive correlations between partons at small $x$  which have to be included in the calculation of the DPS cross section. These soft correlations can be calculated using the connection between MPI and inelastic diffraction.

This
non-factorized contribution to $_2$GPD  is calculated at the initial scale $Q_0^2$ that separates soft and hard physics and which we consider as the starting scale for the DGLAP evolution.
  One expects that for this scale
the single parton distributions at small $x$ are given by the soft Pomeron  and soft Reggeon exchange.

 \begin{figure}[h]  %  figure placement: here, top, bottom, or page
  % \centering
%  \vspace*{-0.3cm}
%   \hspace{0.1cm}
\includegraphics[width=0.48\textwidth]{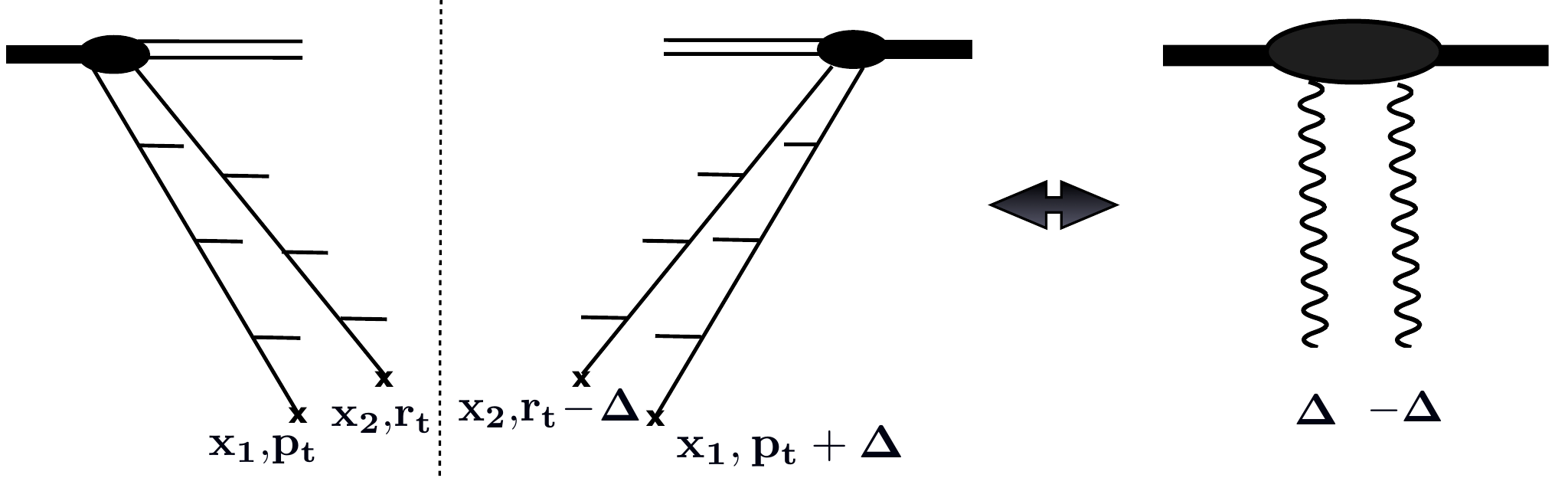}
   \caption{ $_2$GPD as a two Pomeron exchange}
    \label{geom1}
 \end{figure}

\begin{figure}[h]  %  figure placement: here, top, bottom, or page
  % \centering
%  \vspace*{-0.3cm}
  \includegraphics[width=0.48\textwidth]{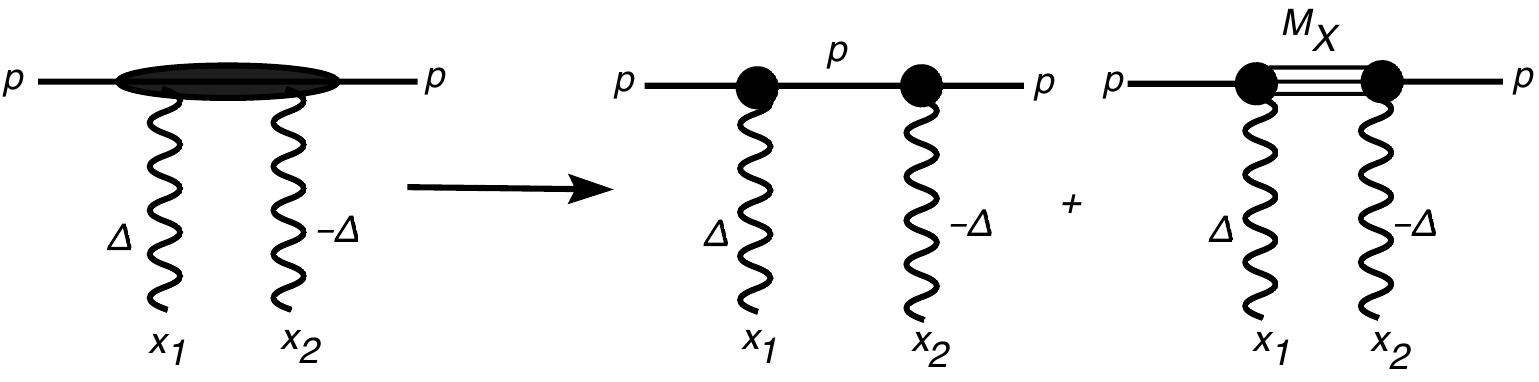}
  \vspace*{-0.3cm}
   \caption{$2\Pomeron$
   contribution to $_2$D and Reggeon diagrams}
    \label{geom2}
 \end{figure}
The diagrams of Fig.~\ref{geom1},\ref{geom2} lead to a simple expression for the non-factorizable/correlated  contribution.
(see \cite{BS1} for details).
For the correlated contribution we have,
\begin{eqnarray}
_2D(x_1,x_2,Q_0^2)_{nf}=c_{3\Pomeron}\int^1_{x_m/a} \frac{dx}{x^2}D(x_1/x,Q_0^2)D(x_2/x,Q_0^2)(\frac{1}{x})^{\alpha_{\Pomeron}}.\nonumber\\[10pt]
+c_{\Pomeron\Pomeron\R}\int^1_{x_m/a} \frac{dx}{x^2}D(x_1/x,Q_0^2)D(x_2/x,Q_0^2)(\frac{1}{x})^{\alpha_{\rm \R}}\nonumber\\[10pt]
\label{d1}
\end{eqnarray}
Here  $x_m=max(x_1,x_2)$.  We also introduced an additional factor  of $a=0.1$ in the limit of integration over $x$ (or, equivalently, the limit of integration over diffraction masses $M^2$) to take into account that the Pomeron exchanges  should occupy  at least two  units in rapidity, i.e. $M^2<0.1\cdot min(s_1,s_2)$ ($s_{1,2}=m_0^2/x_{1,2}$), or $x>max(x_1,x_2)/0.1$, where
 $m_0^2=m_N^2=1$ GeV$^2$ is the low limit of integration over diffraction masses.  Here $c_{3\Pomeron}$ and
$c_{\Pomeron\Pomeron\R}$ are normalized three Pomeron and Pomeron-Pomeron-Reggeon vertices.
We determine   $c_{3\Pomeron}$ and $c_{\Pomeron\Pomeron\R}$ from the HERA data \cite{H1}
for the   ratio of inelastic and elastic diffraction at $t=0$:
\beq
\omega\equiv { {d\sigma_{in.\,  dif.}\over d t}\over {d\sigma_{el}\over d t}}{\left. \right\vert_{t=0} } =0.25 \pm 0.05,
\eeq
and from analysis of diffraction for large $x$ carried in \cite{Luna}, which shows that $c_{\Pomeron\Pomeron\R}\sim 1.5 c_{3\Pomeron}$

We are considering   here
 relatively low energies
 (relative large $x$)
  and a rather modest energy interval. Hence
   we
   neglect energy dependence of
   $c_{3\Pomeron}$.
%%%%%%%%%%%%%%%%%%%%%%%%%%%%%%%%%%%%%%%%%%%%%%%%%%%%%%%%%%%%%%%%%%%%%%%%%%%%%%%%%%%%%%%%%%%%%%%%%%%%%%%%%%%%%%%%%%%%%%%%%%%%%%%%
    Numerically, we obtain $c_{3\Pomeron}=0.075\pm 0.015, c_{\Pomeron\Pomeron\R}\sim 0.11\pm 0.03$ for $Q_0^2=0.5 $ GeV$^2$
    and $c_{3\Pomeron }=0.08\pm 0.015$ and $c_{\Pomeron\Pomeron\R}=0.12\pm 0.03$ for $Q_0^2=1. $ GeV$^2$ ,using the Pomeron intercept values given below.

%%%%%%%%%%%%%%%%%%%%%%%%%%%%%%%%%%%%%%%%%%%%%%%%%%%%%%%%%%%%%%%%%%%%
Note that the intercept of the Pomeron that splits into 2 (region between 2 blobs in fig. 3) is always 1.1 for $t=0$, i.e. this  Pomeron
is by definition soft, and the intercept of Reggeon is 0.5.
\par Note that in \cite{BS1} we used the model that contained only dominant Pomeron component. Here we need
to consider both Pomeron and Reggeon contributions, since we need to go to larger $x$ and smaller diffractive masses. Hence we need to
take into account the
Reggeon contribution which gives dominant contribution to  diffraction for small diffractive masses. Hence the value of $c_{3\Pomeron}$ here is slightly different from the one in \cite{BS1}. However in the kinematics we are interested in,
 once the value of $\omega$ is fixed, the numerical results depend only weakly
on $c_{3\Pomeron}/c_{\Pomeron\Pomeron \R}$ ratio.
\par  For the parton  density in the ladder we use
 \cite{BS1}:
\beq
xD(x,Q_0^2)=\frac{1-x}{x^{\lambda (Q_0^2)}},
\eeq
where the small x intercept of the parton density $\lambda$ is taken from
 the GRV parametrization \cite{GRV98} for the nucleon gluon  pdf at $Q_0^2$ at small x. Numerically $\lambda(0.5 {\rm GeV}^2)\sim 0.27$,
 $\lambda(1.0 {\rm GeV}^2)\sim 0.31$.
%Also, we introduce a cutoff  $M^2/s \le 0.1$ in eq.~\ref{d1} to separate the diffractive contribution (numerically we are not %sensitive to this cutoff in the region where discussed non-factorized contribution is significant).

\par Consider now the $t= - \Delta^2$ dependence of the above expressions.
 The t-dependence of elastic diffraction is
given by
\beq
 F(t)=F_{2g}^2(x_1,t)=\exp (B_{\rm el}(x_1)t).
\eeq

Thus  the t dependence of the  factorized contribution to $_2D_f$ is given by
\beq F(t)=F_{2g}(x_1,t)\cdot F_{2g}(x_2,t)=\exp((B_{\rm el}(x_1)+B_{\rm el}(x_2))t/2),\eeq
where $F_{2g}$ is the two gluon nucleon form factor.
\par The t-dependence of the non-factorized term eq. \ref{d1}
is given by the t-dependence of the inelastic diffraction: $\exp(B_{\rm in}t)$.
Using the  exponential parameterization $\exp(B_{\rm in}t)$ for the t-dependence of the square of the {\em inelastic vertex}\/ $pM_X\!\Pomeron$,
the experimentally measured ratio of the slopes $B_{\rm in}/B_{\rm el}  \simeq 0.28$ \cite{Aaron:2009xp}
translates into the absolute value $B_{\rm in} = 1.4 \div 1.7\, {\rm GeV}^2$.
\par The evolution of the  initial conditions, eq.~ \ref{d1}, is given  by
 \beq
 _2D(x_1,x_2,Q_1^2,Q_2^2)_{\rm nf}=\int_{x_1}^{1}\frac{dz_1}{z_1}\int_{x_2}^{1}\frac{dz_2}{z_2}G(x_1/z_1,Q_1^2,Q_0^2)G(x_2/z_2,Q_2^2,Q_0^2) _2D(z_1,z_2,Q_0^2)_{\rm nf}, \label{r}
 \eeq
  where $G(x_1/z_1,Q_1^2,Q_0^2)$ is the conventional DGLAP gluon-gluon kernel \cite{DDT} which
  describes evolution from $Q_0^2$ to $Q_1^2$.
In our calculation we neglect initial sea quark
 densities in the Pomeron at scale $Q_0^2$ (obviously Pomeron does not get contribution from the valence quarks).

\par Numerical calculation of this integral
\beq
K(x_1,x_2,Q_1^2,Q_2^2)\equiv \frac{D(x_1,x_2,Q_1^2,Q_2^2,Q_0^2)_{\rm nf}}{D(x_1,Q_1^2)D(x_2,Q_2^2)}.
\eeq
 for $Q_0^2=0.5$ GeV$^2$ and
$Q_0^2=1.0 $ GeV$^2$ is presented  in Fig.4.
(the corresponding $x_i=\sqrt{4Q^2/s}$ ).
 \begin{figure}[htbp]
\begin{center}
\includegraphics[scale=0.7]{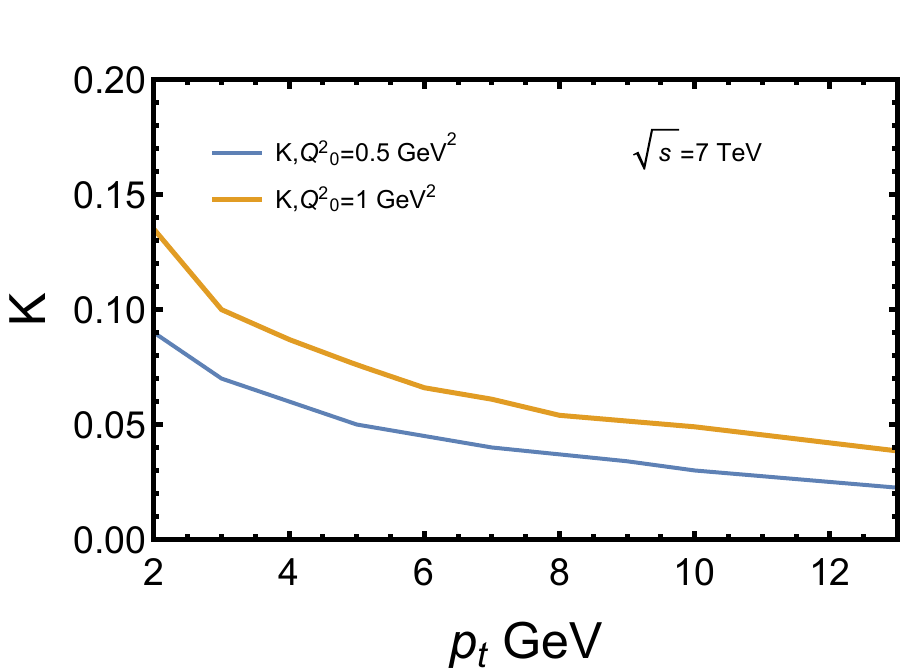}\hspace{3cm}
\includegraphics[scale=0.7]{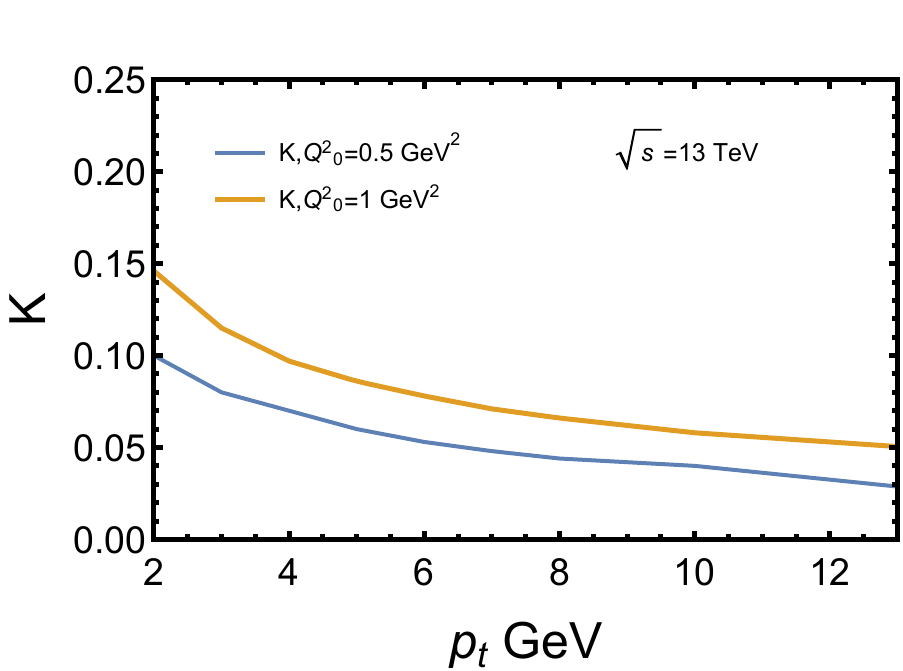}\hspace{3cm}
\label{c2}
\caption{The ratio of non-factorized and factorized contributions to $_2D$, K(t=0), as a function of transverse scale $p_t\equiv Q$ in maximum transverse kinematics for $\sqrt{s}$= 7 TeV run (left) and $\sqrt{s}$= 13.0 TeV run (right)}
\end{center}
\end{figure}

\section{\effs \, in the central kinematics}.
\par Now we are in the position to determine the overall enhancement of $1/ \sigma_{\rm eff}$
as compared to the mean field result. It is
 given by the
enhancement coefficient \beq
R=R_{pQCD}+R_{soft}.
\eeq
Here as it was already explained in section 2,  $R_{pQCD}$ corresponds to the contribution of \12 pQCD mechanism (Fig. 1b) and was calculated in \cite{BDFS4}.
The
 expression for $R_{soft}$ is given by
\beq
R_{soft}=\frac{4K}{1+B_{inel}/B_{el}}+\frac{K^2B_{el}}{B_{in}}+KR_{pQCD}B_{el}/B_{inel},
\eeq
where  we calculate all factors for $x_1=x_2=x_3=x_4=\sqrt{4Q^2/s}$, with $s$ being invariant
energy of the collision.
We present our numerical results in Figs. 5 -- 8.
 In Fig. 5,6  we present  \effs as a function of $p_t$ for two values of the starting evolution scale for the central kinematics in the mean field approach, accounting also for the pQCD \12 mechanism and including in addition soft correlations.

In fig. 7,8 we show the corresponding enhancement
 factors.
 \begin{figurehere}
\begin{center}
\includegraphics[scale=0.7]{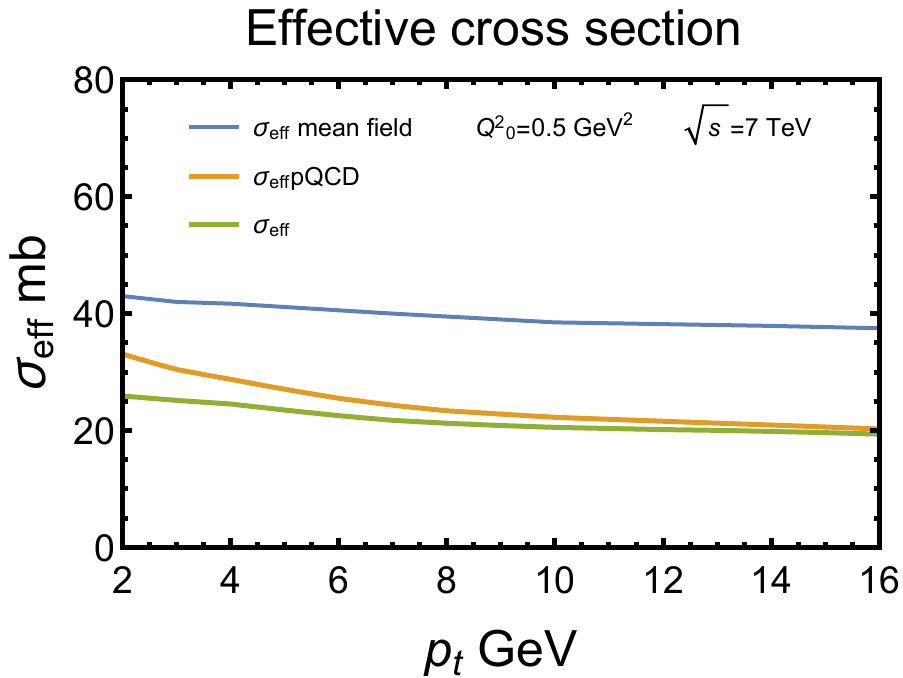}\hspace{3cm}
\includegraphics[scale=0.7]{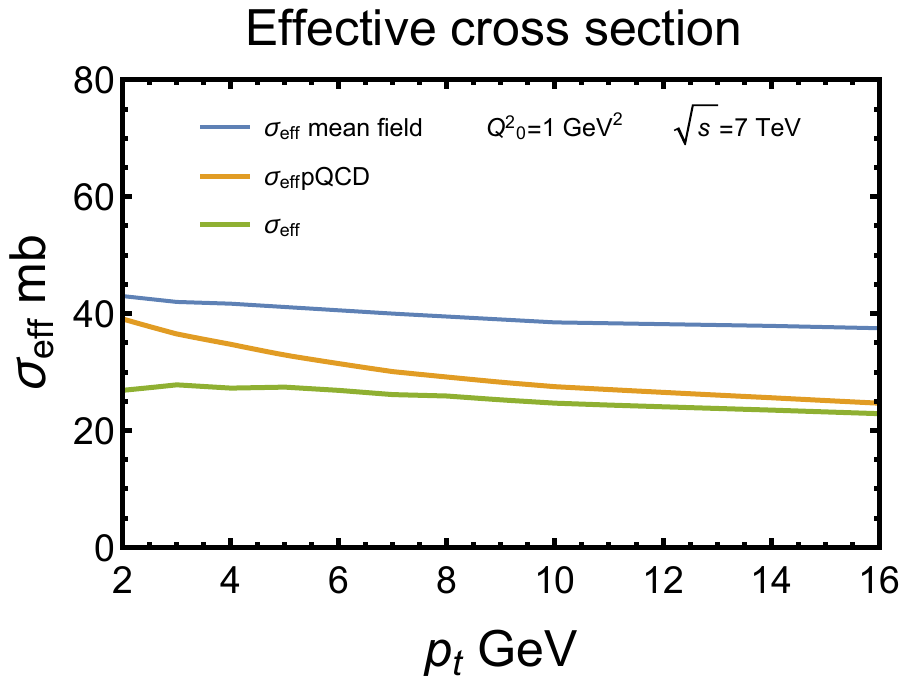}\hspace{3cm}
\label{5b3}
\caption{\effs \,  as a function of the  transverse scale $p_t$ for $Q_0^2=0.5$ (left),$1$ GeV$^2$  (right)in the central kinematics.
We present the mean field, the mean field plus \12 mechanism and total \effs  for $\sqrt{s} $=  7 TeV.}
\end{center}
\end{figurehere}
 \begin{figurehere}
\begin{center}
\includegraphics[scale=0.7]{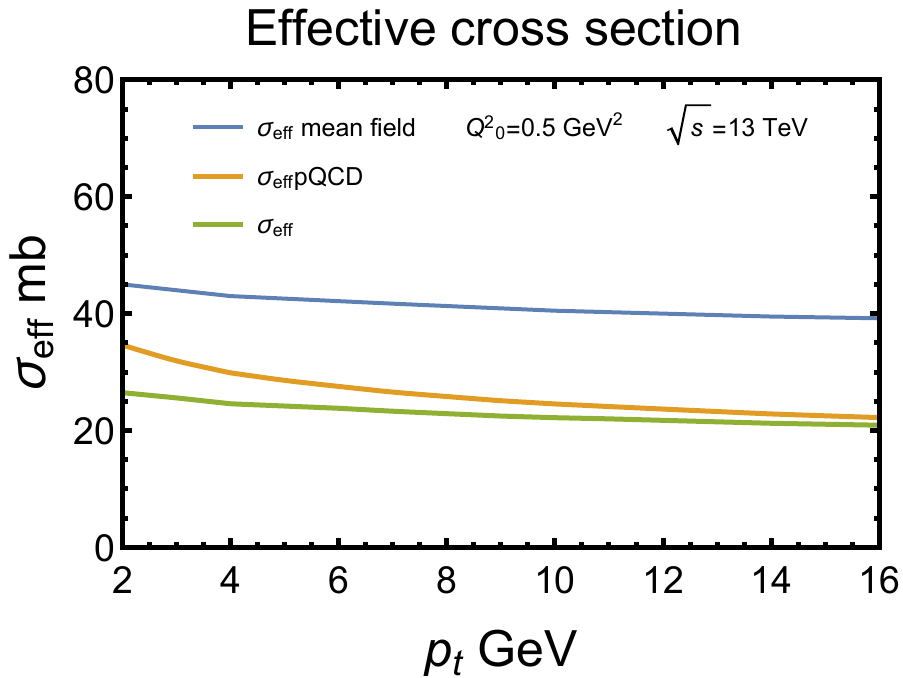}\hspace{3cm}
\includegraphics[scale=0.7]{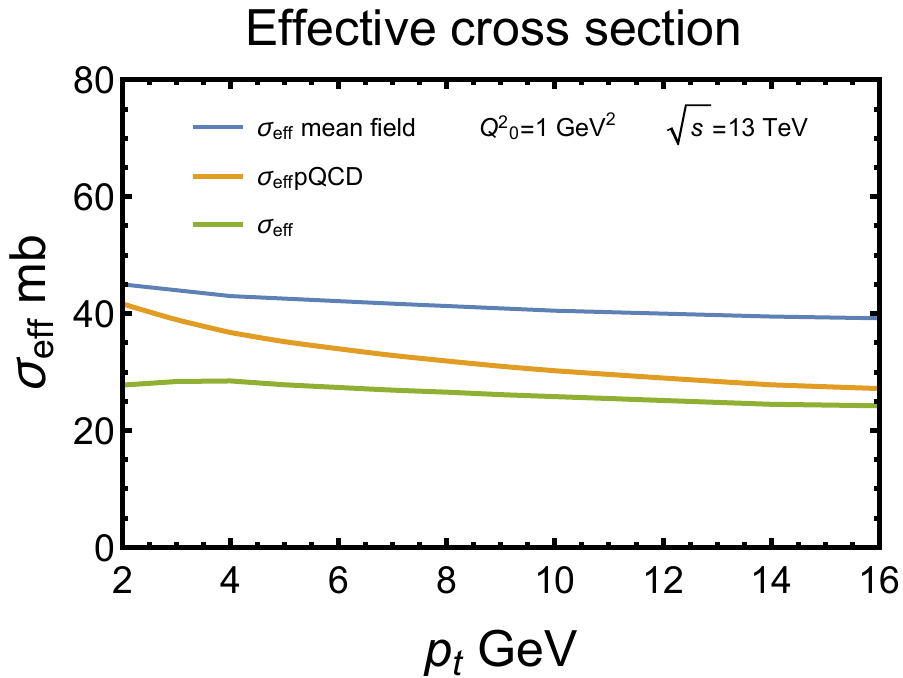}\hspace{3cm}
\label{5b4}
\caption{\effs \,  as a function of the  transverse scale $p_t$ for $Q_0^2=0.5$ (left),$1$ GeV$^2$  (right)in the central kinematics.
We present the mean field, the mean field plus \12 mechanism and total \effs  for  $\sqrt{s} $=  13 TeV.}
\end{center}
\end{figurehere}

\begin{figurehere}
\begin{center}
\includegraphics[scale=0.7]{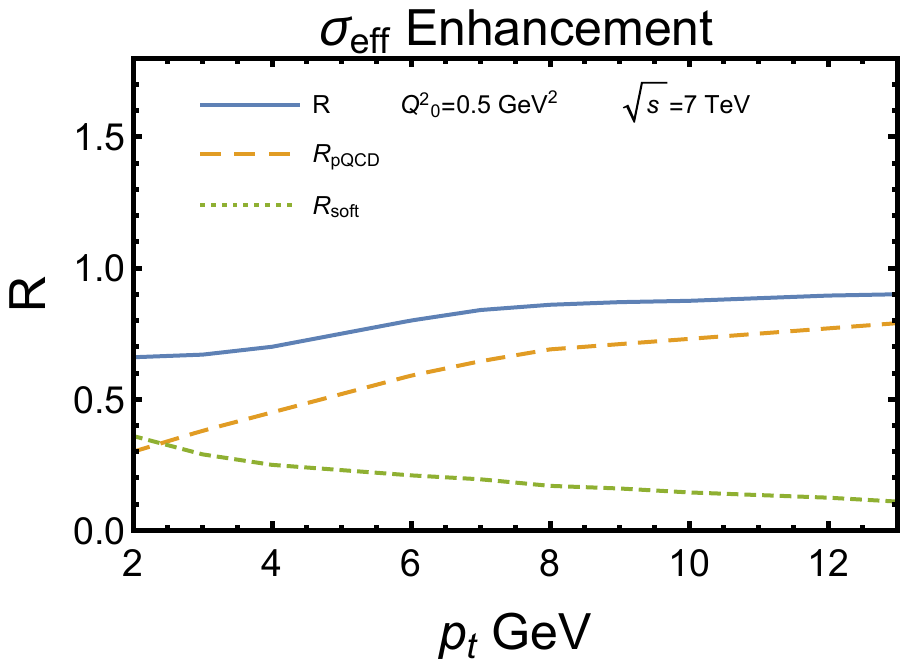}\hspace{3cm}
\includegraphics[scale=0.7]{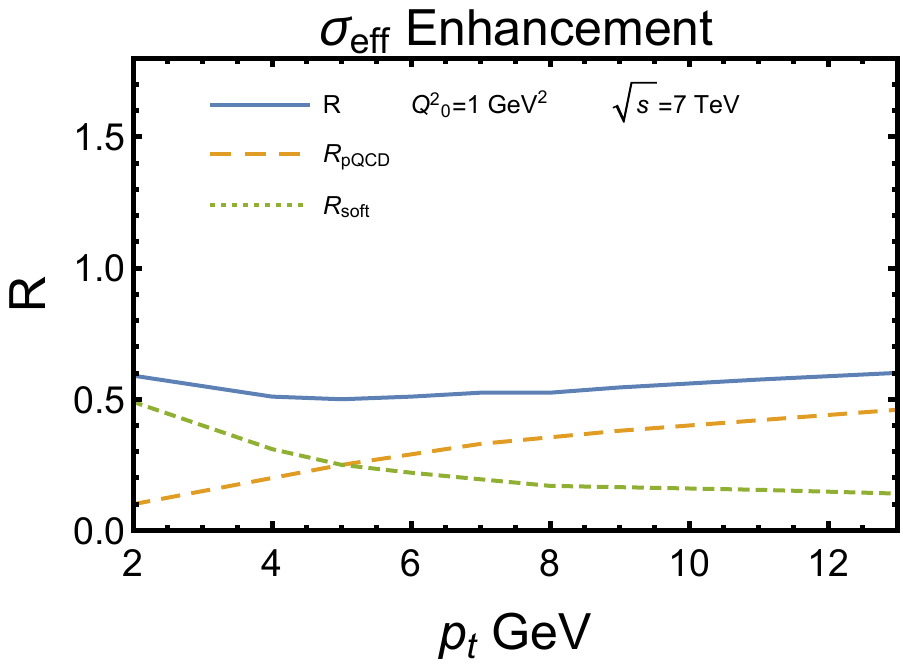}\hspace{3cm}
\label{5b1}
\caption{R for different $Q_0^2$ and $\sqrt{s} $ =7 TeV.}
\end{center}
\end{figurehere}
\begin{figurehere}
\begin{center}
\includegraphics[scale=0.7]{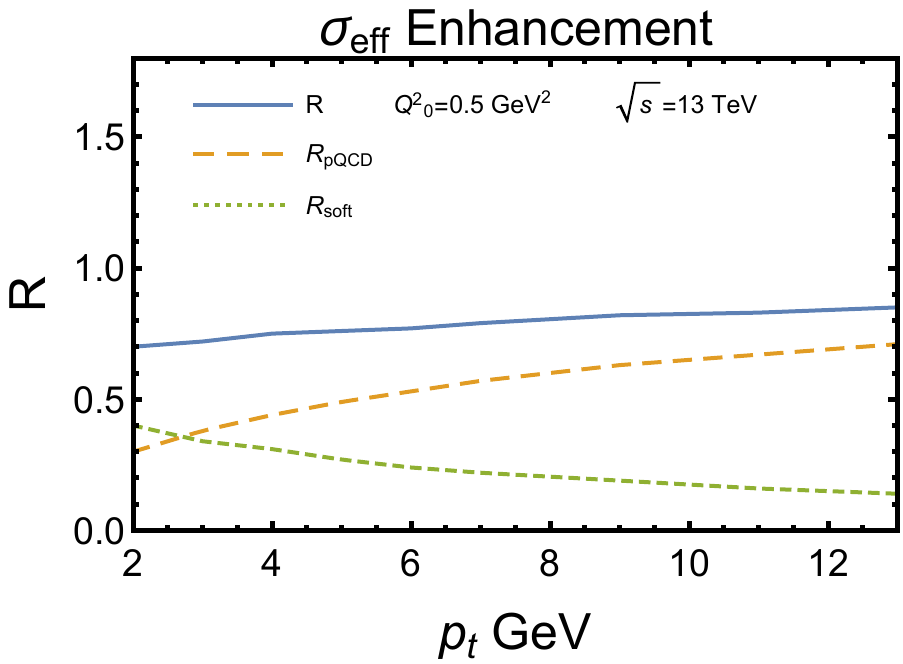}\hspace{3cm}
\includegraphics[scale=0.7]{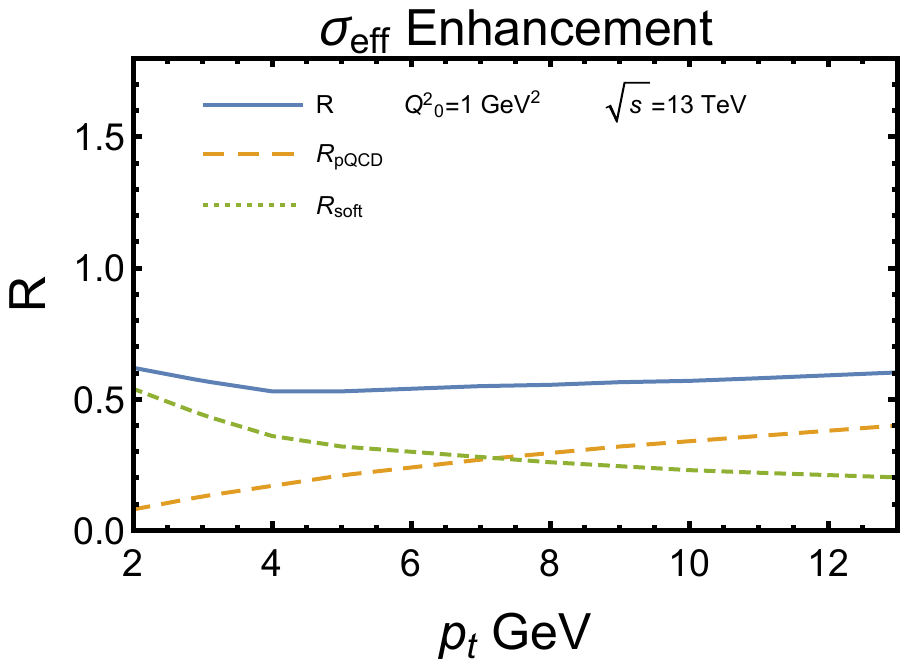}\hspace{3cm}
\label{5b2}
\caption{R for different $Q_0^2$ and $\sqrt{s}$ =13 TeV.}
\end{center}
\end{figurehere}
\par In addition, in order to
%envision
illustrate
 the combined picture of \effs behaviour in both UE and DPS, in Fig.9 we give the example
of \effs behaviour as a function of $p_t$ in the combined transverse momenta region 2-50 GeV for $Q_0^2$=0.5 GeV$^2$ (for $Q_0^2=1 $ GeV$^2$
the behaviour is very similar).
\begin{figurehere}
\begin{center}
\hspace{2cm}
\mbox{
\includegraphics[scale=0.7]{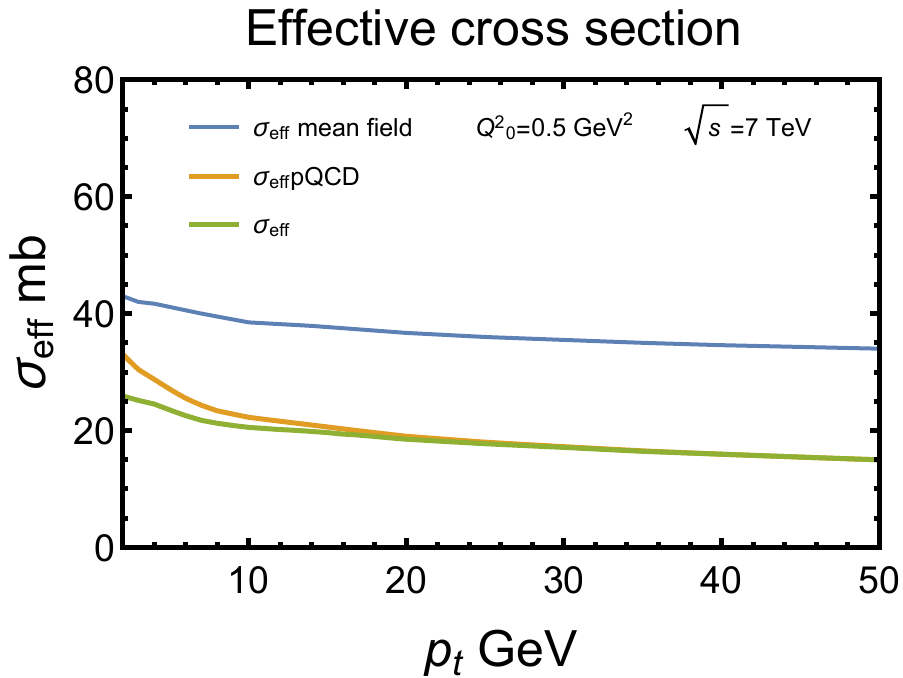}}\hspace{3cm}
\label{5b7}
\caption{\effs for entire transverse momenta region ($Q_0^2=0.5$ GeV$^2$)}
\end{center}
\end{figurehere}
\par We also studied the energy dependence of \effs for fixed transverse momenta $p_t$ on center of mass energy in the UE kinematic region in the energy region from Tevatron to LHC, that is depicted in Fig. 10:

\begin{figurehere}
\begin{center}
\hspace{2cm}
\mbox{
\includegraphics[scale=0.7]{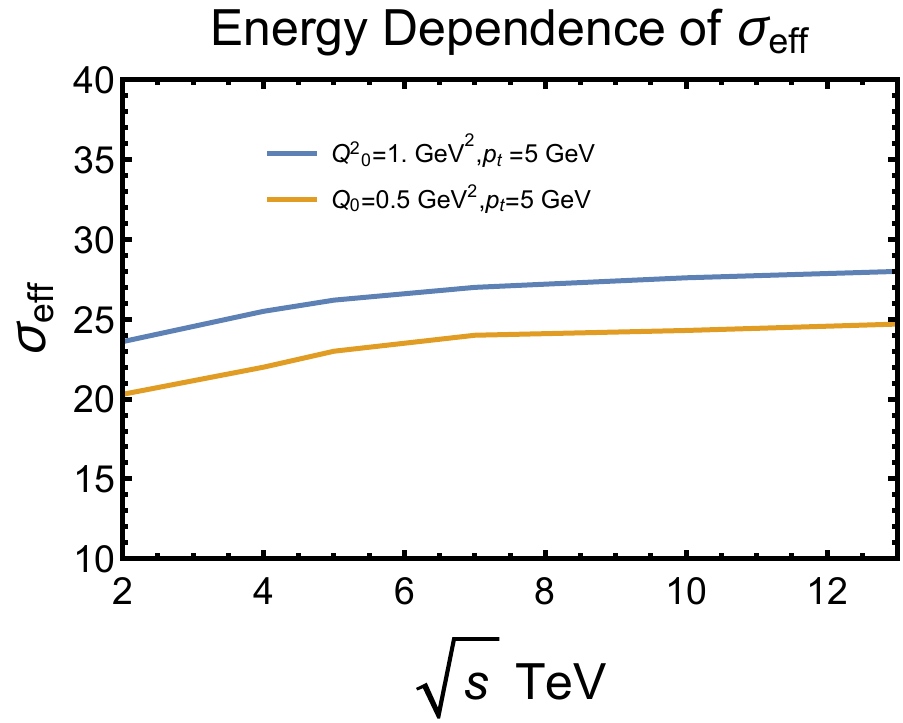}}\hspace{3cm}
\label{5b8}
\caption{The characteristic energy dependence of \effs on c.m.s. energy $\sqrt{s}$}
\end{center}
\end{figurehere}

We see that at LHC energies the energy dependence of \effs practically saturates. In order to understand
the evolution of \effs for higher energies for given transverse scale we shall need the evolution of two-gluon formfactor
for small x$\le 10^{-4}$ which is still not available from experimental data.

\section{Conclusions}
\par We
used the model of \cite{BS1} to study nonperturbative parton - parton correlations in the central kinematics at the LHC.
Our estimates have  been only
semiquantitative due to the
 large uncertainties in diffraction parameters as well us the use of the
"effective" values for the reggeon/pomeron parameters (which very roughly included
screening corrections). Nevertheless,
our results indicate
a number of basic features of soft nonperturbative parton - parton correlations relevant for the
 central LHC dynamics, relevant for ATLAS, CMS and ALICE detectors.

\par (i)  We see that for large transverse momenta, relevant for hard DPS scattering, soft effects are small and essentially negligible, contributing only $5\%$ to the enhancement coefficient R if we start from the scale $Q_0^2=0.5$ GeV$^2$, and $10-15\%$ from $1$ GeV$^2$, for $p_t\sim 15-20$ GeV.
Thus they do not influence detailed hard DPS studies carried in \cite{BDFS3,Gauntnew}.
Our results also indicate that the characteristic transverse momentum $p_{t0}$, for  which soft correlations constitute given fixed
part of enhancement R rapidly increase with s.
Indeed, if we look at the scale where soft contribution are say $10\%$ of the rescaling coefficient R
changes from 12 to 15 GeV once we change c.m.s. energy from $\sqrt{s_1}=7$ to $\sqrt{s_2}=13$ TeV. This scale further changes to $\sim 6$ GeV
once we go to 2 TeV c.m.s. energy corresponding to Tevatron. The  dependence of $p_{t0}$ on energy starting from Tevatron energies is depicted in Fig.11 for $Q_0^2=0.5$ GeV$^2$ (for $Q_0^2=1$ GeV$^2$ the qualitative behaviour of $p_{t0}$ is rather similar).
We see that the characteristic transverse
scales for which soft correlations are important, rapidly increase with energy.
\begin{figurehere}\begin{center}
\hspace{3cm}
\mbox{\includegraphics[scale=0.5]{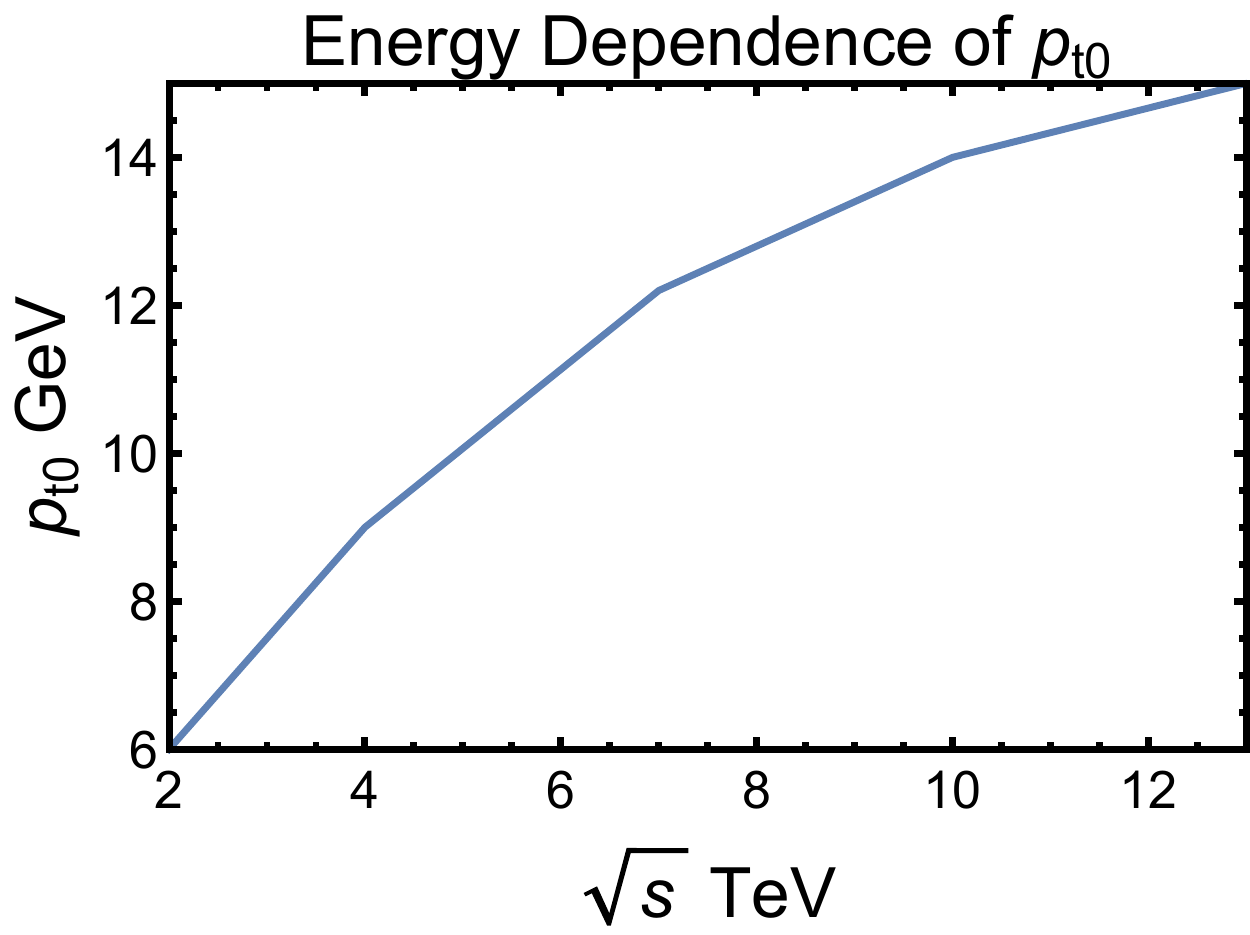}}
\hspace{3cm}
\label{5b9}
\caption{The characteristic energy dependence of $p_{t0}$ on c.m.s. energy $\sqrt{s}$}
\end{center}
\end{figurehere}
\par (ii)  The soft non-factorisable contributions may contribute significantly in the underlying event dynamics, especially
at the scales 2-4 GeV where they are responsible for about 50\% of
the difference between mean field result and full prediction for
$\sigma_{\rm eff}$ for 0.5 GeV$^2$ case, and are dominant up to 4 GeV if we start evolution from 1 GeV$^2$.
In UE they lead to stabilization of \effs, that decreases more slowly with increase of $p_t$ than if we include only perturbative correlations,i.e. \12 mechanism. For UE dynamics  if we use 0.5 GeV$^2$ as a start of the evolution, and take $p_t\sim 2 $ GeV,
\effs changes from 43mb (mean field value) to 26 mb, while if we neglect soft correlations the change from mean field value
is to 33 mb (for $\sqrt{s}=7$ TeV).
 Similarly if we start evolution at  $Q_0^2=1$ GeV$^2$, the change is to 27 mb, instead of 39 mb, if only pQCD effects are included.
For $p_t\sim 4$ GeV the changes are from 42 mb, to  26-27 mb (instead of 30-35 mb if soft correlations are not accounted for).
  These values for \effs for UE, especially for scales 2-4 GeV are very close to the ones used by Pythia.
The results show that \effs in UE significantly decreases as compared to the values one obtains   including
only contributions of the  mean field and \12 mechanism.
Thus for  the 2-4 GeV scales, the full value
of \effs that includes soft correlations
 is 26-27 mb, instead of 30-39 mb that we find  with only pQCD and mean field contributions included.
\par Note that the pQCD contribution $R_{pQCD}$ was included in the MC calculations in
\cite{BG1}. It was shown that its inclusion leads to a significant improvement of the description of DPS scattering
for $p_t>15-20$ GeV. On the other hand for UE regime the best agreement was for  the Pythia nonrescaled tune.
Our current results for the Underlying event are very   close to Pythia, while for DPS with $p_t \ge  \mbox{15-20 GeV}$ effectively
 only mean field+\12 pQCD terms contribute.
 We see that the new framework can  give a decent description of the data over the full transverse momenta range, but with less dependence of the quality of fit on the starting point of the evolution $Q_0$ than in \cite{BG1}. (Of course more detailed comparisons of the old and new frameworks will be ultimately needed. Such more detailed MC simulations and comparison with experimental data are  currently under %investigation
 way
 \cite{BG3}.)
\par (iii) The evolution of \effs with transverse scale stabilizes for UE regime, as it is seen in Fig. 5,6 leading to almost plateau like picture with only slight decrease with transverse scale.
\par  (iv) The inclusion of soft correlations  stabilizes the energy, i.e. $\sqrt{s}$ dependence of  \effs . It changes marginally
 between 3.5 TeV and 6.5 TeV collision energies  for the same transverse scale for small $p_t$. In other words the increase of soft correlations compensates the decrease of the relative pQCD contribution with an increase of energy (i.e. decrease of effective $x_i$ in the process).

\par Overall we conclude
  that soft correlations do not influence significantly hard DPS dynamics, but are important for the description of MPI in the UE regime.
\par Note also that inclusion of the  soft correlations decreases the difference between $R$ obtained while carrying evolution
from $Q_0^2=0.5$ and $Q_0^2=1$ GeV$^2$ scales, especially for small transverse scales of several GeV.
\par Our results do not influence our
previous
results for DPS scattering in the Tevatron \cite{BDFS4}, where soft correlations give only $2-5\%$
contribution, since the corresponding region
of $x$ is $x_i\sim 0.01$. This is in accordance with increase of characteristic momenta where soft correlations cease to be important with energy mentioned above.
\par Finally, we note that an attempt to include soft correlations, which is also based on the ideas of ref. [9], was recently reported  [36]. There are several important differences in the used models:
 \par (a) in [36] scale $Q/2$ was used  for production of jets rather more commonly used scale $Q$,
 \par (b) the mean field  contribution was calculated using
 the  two gluon form factor  with the t-dependence much harder that the one allowed by the exclusive $J/\psi$ data; (c) In [36] it was assumed
 that the QCD evolution starts only at $Q^2 \sim 3 \mbox{GeV}^2$.
Combined these assumptions resulted in an enhancement of nonperturbative contribution to \effs
 and suppression of the perturbative contribution.
\par (c) Another difference is that
we used the  effective values for Reggeon  parameters
in the spirit of estimates in ref. [31], and GRV gluon densities for the Pomeron, while the
authors of [36]  used a version of the full Reggeon Field theory with screenings. Note in passing   that the Reggeon Field theory was formulated by Gribov assuming presence of only one (soft) scale   in strong interaction. Such an assumption is difficult to justify for the LHC energies where minijet  cross section is very large.
\par
Numerically we find that for the UE our numerical results are rather close (see Fig. 10 and corresponding figure in [36].)

While the authors of ref. [36] find \effs =  25$\div$ 25.5 mb
for energy range considered in our  letter,  we get, depending on $Q_0$, \effs $ \sim 20\div 25$ mb for $Q_0^2=0.5$ GeV$^2$, and 24$\div$ 28 mb for
$Q_0^2=1$ GeV$^2$, i.e. our results differ by 20 $\%$ or less.
The pattern of slow increase of \effs with energy and its almost complete saturation for the LHC highest energies are also similar. Such similarity appears rather
accidental,
 as it results from very different model for the gluon GPD used for the mean field, and different
assumptions about the range of QCD evolution.
\par Indeed, for large scale considered in [36] of order 50 GeV we have a value of \effs approximately 1.5-1.7 times smaller
[10] than in [36], and for such scales soft correlation contribution is negligible.
\par Note also that  we do not extend our calculation of UE beyond the  LHC
energies, as it was done in ref.[36], since the dependence on $x$ of the two gluon form factor  and inelastic diffraction on $x$ are not known so far for $x < 10^{-4}$.
\section*{Acknowledgements}
\label{Ack}
M.S.'s research was supported by the US Department of Energy Office of Science,
Office of Nuclear Physics under Award No.  DE-FG02-93ER40771.
We thank  Yuri Dokshitzer and Leonid Frankfurt for many useful discussions.

\end{document}